\documentclass[showpacs,twocolumn,aps,prl,floatfix,superscriptaddress,10pt]{revtex4-1}
\pdfoutput=1
\usepackage{amssymb}
\usepackage{amsmath}
\usepackage{amsfonts}
\usepackage{graphicx}

\newcommand{\ket}[1]{|#1\rangle}
\newcommand{\bra}[1]{\langle #1|}

\begin{document}

\title{Quenching of the quantum Hall effect in graphene with scrolled edges}

\author{Alessandro Cresti}
\affiliation{IMEP-LAHC (UMR CNRS/INPG/UJF 5130), Grenoble INP, Minatec,
3 Parvis Louis N\'{e}el, BP 257, F-38016 Grenoble, France}

\author{Michael M. Fogler}
\affiliation{Department of Physics, University of California San Diego, 9500 Gilman Drive, La Jolla, California 92093, USA}

\author{Francisco Guinea}
\affiliation{Instituto de Ciencia de Materiales de Madrid, CSIC, Cantoblanco, E-28049 Madrid, Spain}

\author{A. H. Castro Neto}
\affiliation{Graphene Research Centre and Department of Physics, National University of Singapore, 2 Science Drive 3, 117542, Singapore}

\author{Stephan Roche}
\affiliation{CIN2 (ICN-CSIC) and Universitat Autonoma de Barcelona, Catalan Institute of Nanotechnology, Campus de la UAB, 08193 Bellaterra (Barcelona), Spain}
\affiliation{ICREA, Institucio Catalana de Recerca i Estudis
Avan\c{c}ats, 08010 Barcelona, Spain}

\date{\today}

\begin{abstract}

Edge nanoscrolls are shown to strongly influence transport properties of suspended graphene in the quantum Hall regime. 
The relatively long arc length of the scrolls in combination with their compact transverse size results in formation of many nonchiral transport channels in the scrolls. 
They short-circuit the bulk current paths and inhibit the observation of the quantized two-terminal resistance. 
Unlike competing theoretical proposals, this mechanism of disrupting the Hall quantization in suspended graphene is not caused by ill-chosen placement of the contacts, singular elastic strains, or a small sample size.

\end{abstract}

%72.80.Vp   Electronic transport in graphene
%73.22.Pr   Electronic structure of graphene

\maketitle

%%%%%%%%%%%%%%%%%%%%%%%%%%%%%%%%%%%%%%%%%%%%%%%%%%%%%%%%%%%%%%%%%%%%%%%%%

Graphene is a material that combines a highly tunable metallic conduction~\cite{Netal04, Netal05b} with a flexibility of a two-dimensional (2D) membrane~\cite{NGPNG09, AGSci09}. 
The hallmark of a 2D metal is the quantum Hall effect (QHE), which can be observed in graphene on SiO$_2$ substrates at room temperature. 
A more fragile fractional QHE is robust in graphene on BN substrates~\cite{Dean2011mfq}. 
Yet the demonstration~\cite{Du2009fqh, Bolotin2009oot} of the QHE in graphene suspended off a substrate has been difficult and limited largely to two-terminal measurements instead of the orthodox four-terminal ones. 
In this paper we propose an explanation for this surprising result.

Being a topological property, the quantization of the Hall resistance $\rho_{x y}$ can normally be destroyed only if external perturbations exceed the Landau gap. 
If so, the system can split into regions with different local $\rho_{x y}$ while the observed Hall resistance can have some average non-quantized value. 
However, suspended graphene is believed to have very low disorder~\cite{Meyer2007tso, DSBA08, Betal08a, Booth2008mgm, Betal08b}. 
Another possible reason for deviation from the ideal QHE is backscattering of the chiral edge channels across the sample due to impurities~\cite{Skachko2010fqh} or elastic strains~\cite{Prada2010ses}. 
In samples that are many microns wide this also can be ruled out. The only other known mechanism of disrupting the QHE is an edge reconstruction (ER), which can generate counter-propagating channels at the same edge. 
However, previously discussed ERs~\cite{Chamon1994sas, Wan2003eri, CastroNeto2006eas, Silvestrov2008caa} are not unique to suspended graphene. 
They originate from a generic tendency of a 2D metal to have a nonuniform density near the edge.

%%%%%%%%%%%%%%%%%%%%%%%%%%%%%%%%%%%%%%%%%%%%%%%%%%%%%%%%%%%%%%%%%%%%%%%%%
\begin{figure}
\begin{center}
\includegraphics[width=2.3in]{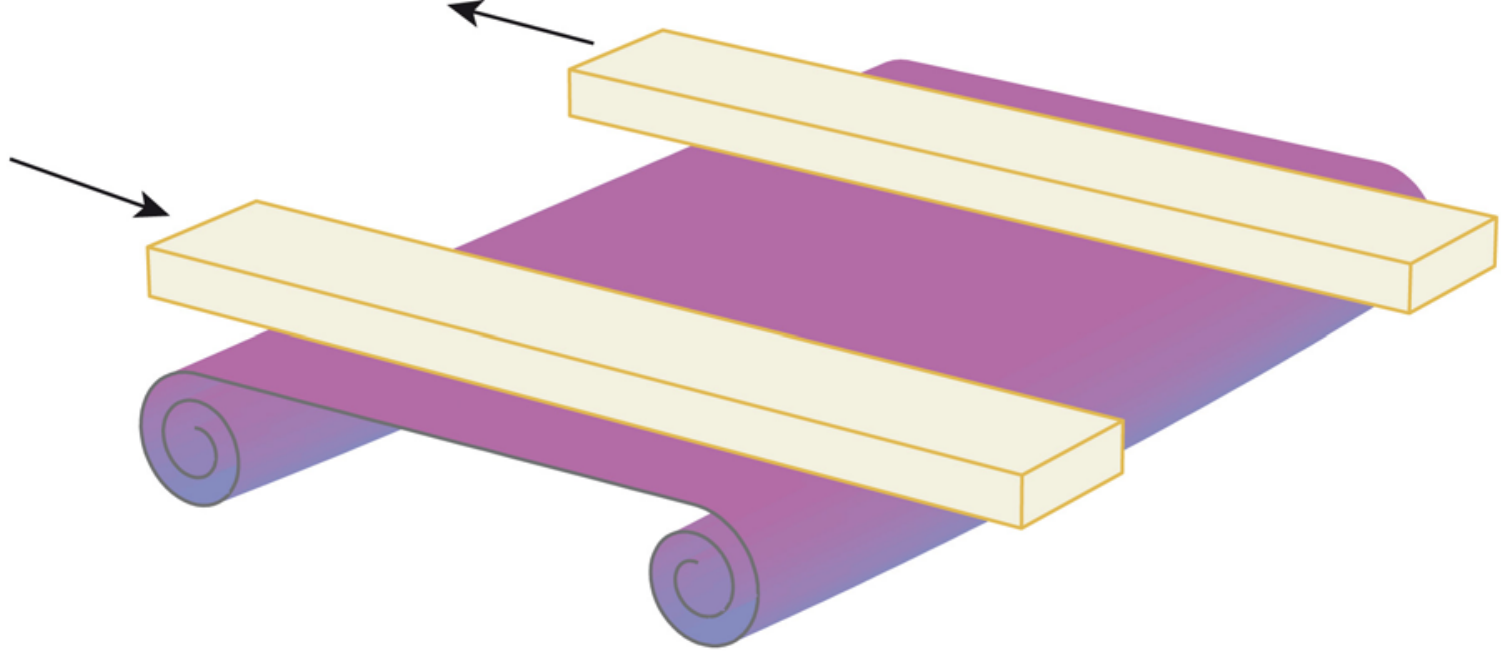}
\caption{\label{fig:scheme}
  (Color online) 
  Sketch of a system under study in which a suspended graphene ribbon (purple) has scrolled edges. 
  The ribbon is attached to the source and drain contacts (gold). The arrows symbolize the external current.}
\end{center}
\end{figure}
%%%%%%%%%%%%%%%%%%%%%%%%%%%%%%%%%%%%%%%%%%%%%%%%%%%%%%%%%%%%%%%%%%%%%%%%%

In this Letter, we examine a novel type of ER  --- edge nanoscrolls --- that is specific and indeed common to free-standing and suspended graphene~\cite{Viculis2003acr, Meyer2007tso, Xie2009cfo, Liu2009oac}. 
Our basic idea is that the magnetic field $B = 2$--$20\,\text{T}$ in typical experiments has only a weak effect on the edge scrolls (if the Fermi energy is large enough). In this respect, the scrolls are similar to multiwall carbon nanotubes (CNT) where the QHE is seen only at very high fields~\cite{[{For review, see }] Charlier2007eat}. 
The reason is that electrons respond primarily to the component $B_\perp$ of the total field along the local normal to the graphene sheet. 
Inside the scrolls, which look like spirals in cross-section (Fig.~\ref{fig:scheme}), $B_\perp$ oscillates in sign and largely averages out. 
In contrast, the flat portion of the sample
is expected to be in the QHE regime and thus has a low dissipative conductivity. 
The current injected from the contacts may prefer to avoid the flat region and instead diffuse into the more conducting scrolls. 
The lack of $\rho_{x y}$ quantization therein would then significantly influence the net two-terminal resistance of the device.

%%%%%%%%%%%%%%%%%%%%%%%%%%%%%%%%%%%%%%%%%%%%%%%%%%%%%%%%%%%%%%%%%%%%%%%%%
\begin{figure}[t]
\begin{center}
\includegraphics[width=2.5in]{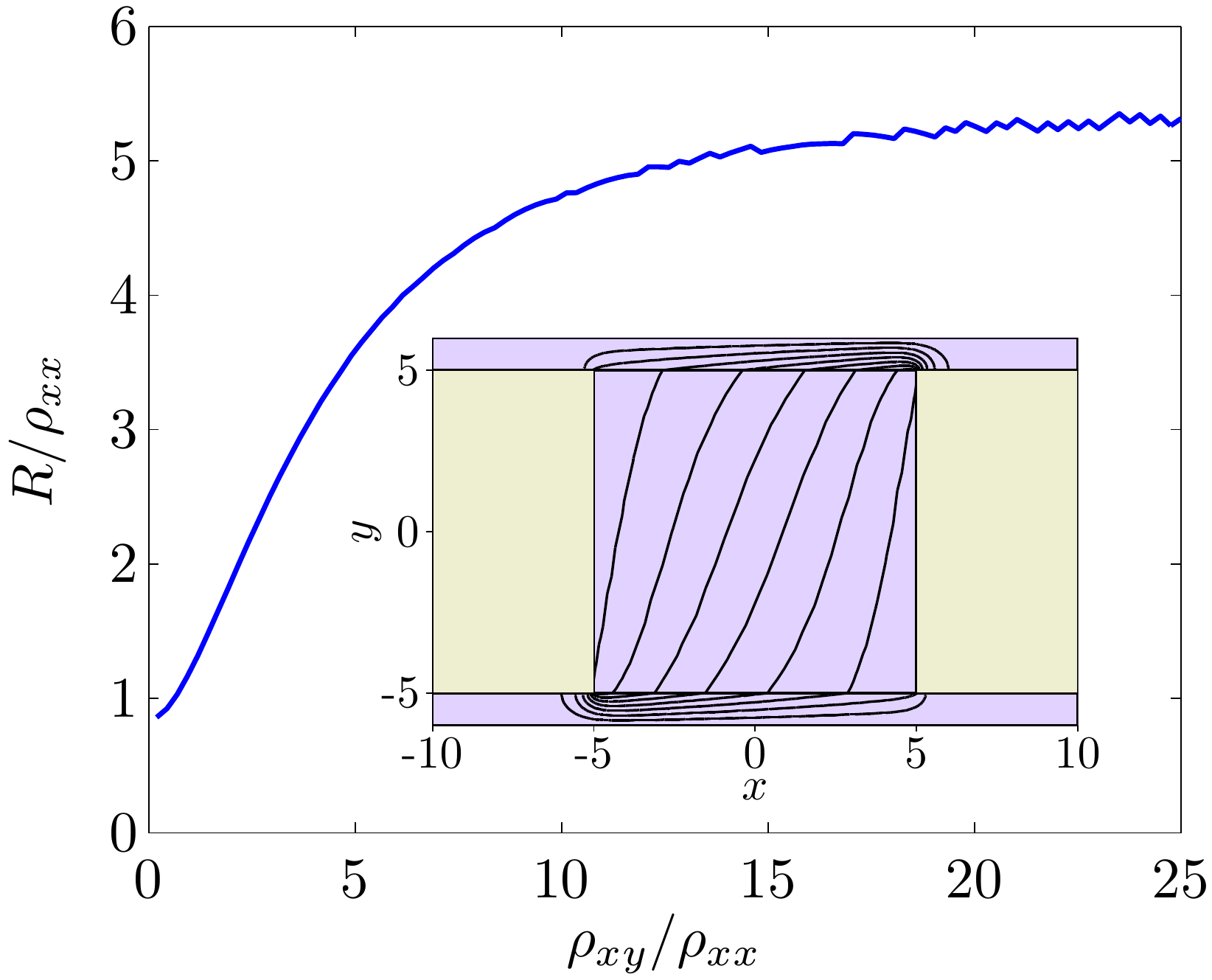}
\caption{\label{fig:current}
  (Color online) 
  Inset: Lines of current in the graphene ribbon (purple) in the semiclassical approximation. For clarity, the edge scrolls are unwrapped and only the parts of the electrodes in contact with the ribbon are shown (gold). 
  Main panel: Two-terminal resistance $R$ as a function of $\rho_{x y}$, both normalized to the constant diagonal resistivity $\rho_{x x}$.}
\end{center}
\end{figure}
%%%%%%%%%%%%%%%%%%%%%%%%%%%%%%%%%%%%%%%%%%%%%%%%%%%%%%%%%%%%%%%%%%%%%%%%%

Below we support this qualitative picture by several calculations, which are carried out at the progressively more microscopic level. 
We start with continuum models suitable for large scrolls. Subsequently, we consider a few nm-wide scrolls using the discrete lattice representation.

By large scrolls, we mean those with outer radius $r_o \sim 10\,\text{nm}$.
Such scrolls may contain many turns and their total arc length can measure in hundreds or thousands of nm~\cite{Xie2009cfo, Liu2009oac, Schaper2011cso, Schaper2011oot}. Adjacent graphene layers of the scrolls may have commensurate~\cite{Chuvilin2009ccn, Ivanovskaya2011} or incommensurate lattice structure. In the latter, more typical, case the scrolls are held together by van~der~Waals attraction~\cite{Tomanek2002mow, Fogler2010eoe} and the separation between the adjacent layers should be about the same as in graphite, $0.34\,\text{nm}$. Assuming that incommensurability makes interlayer electron tunneling negligible, we can switch from the three-dimensional (3D) Fig.~\ref{fig:scheme} to the equivalent 2D Fig.~\ref{fig:current} (inset) of the unwrapped sheet. Discarding the part directly attached (and thus electrically shunted) by the contacts, we obtain the I-shaped silhouette of the unwrapped sample. (Existence of the scrolls underneath the contacts is suggested by published micrographs of suspended graphene~\cite{Elias2011dcr}.) This is the system we want to study the transport through.

We begin with a simple semiclassical approximation in which the transport in graphene is described by local diagonal $\rho_{x x}(x, y)$ and Hall $\rho_{x y}(x, y)$ resistivities. Assuming that magnetic field has no effect on electron states inside the scrolls, we set $\rho_{x y}(x, y)$ to zero in the scrolls and to a nonzero constant in the flat part of the sample. We take $\rho_{x x}$ to be coordinate independent. The current distribution in this model can be easily computed numerically. The lines of the current for some representative parameters are shown in Fig.~\ref{fig:current}~(inset). 
The chosen arc length of each scroll $\ell$ is a small fraction of the width $L = 10 \ell$ of the flat part (the central square). Yet a significant portion of the current flows inside the scrolls. 

The results for the two-terminal resistance $R$ are plotted in the main panel of Fig.~\ref{fig:current}. 
They imply that our system acts as three resistors in parallel. One resistor represents the flat region.
In the QHE regime, where $\rho_{x y} \gg \rho_{x x}$, its magnitude is approximately equal to $R_f = \rho_{x y}$. The other two resistors, of magnitude $R_s = \rho_{x x} L / \ell$ each, represent the scrolls. If $1 \ll \rho_{x y} / \rho_{x x} \ll L / \ell$, which corresponds to weak or modestly strong magnetic fields, the scrolls are not important, so that have $R \simeq R_f \simeq \rho_{x y}$. At larger $\rho_{x y} / \rho_{x x}$, the scrolls short-circuit the transport, so that $R \simeq R_s$, which  is $\rho_{x y}$ independent in this model. 

%%%%%%%%%%%%%%%%%%%%%%%%%%%%%%%%%%%%%%%%%%%%%%%%%%%%%%%%%%%%%%%%%%%%%%%%%
\begin{figure}[b]
\begin{center}
\includegraphics[width=2.7in]{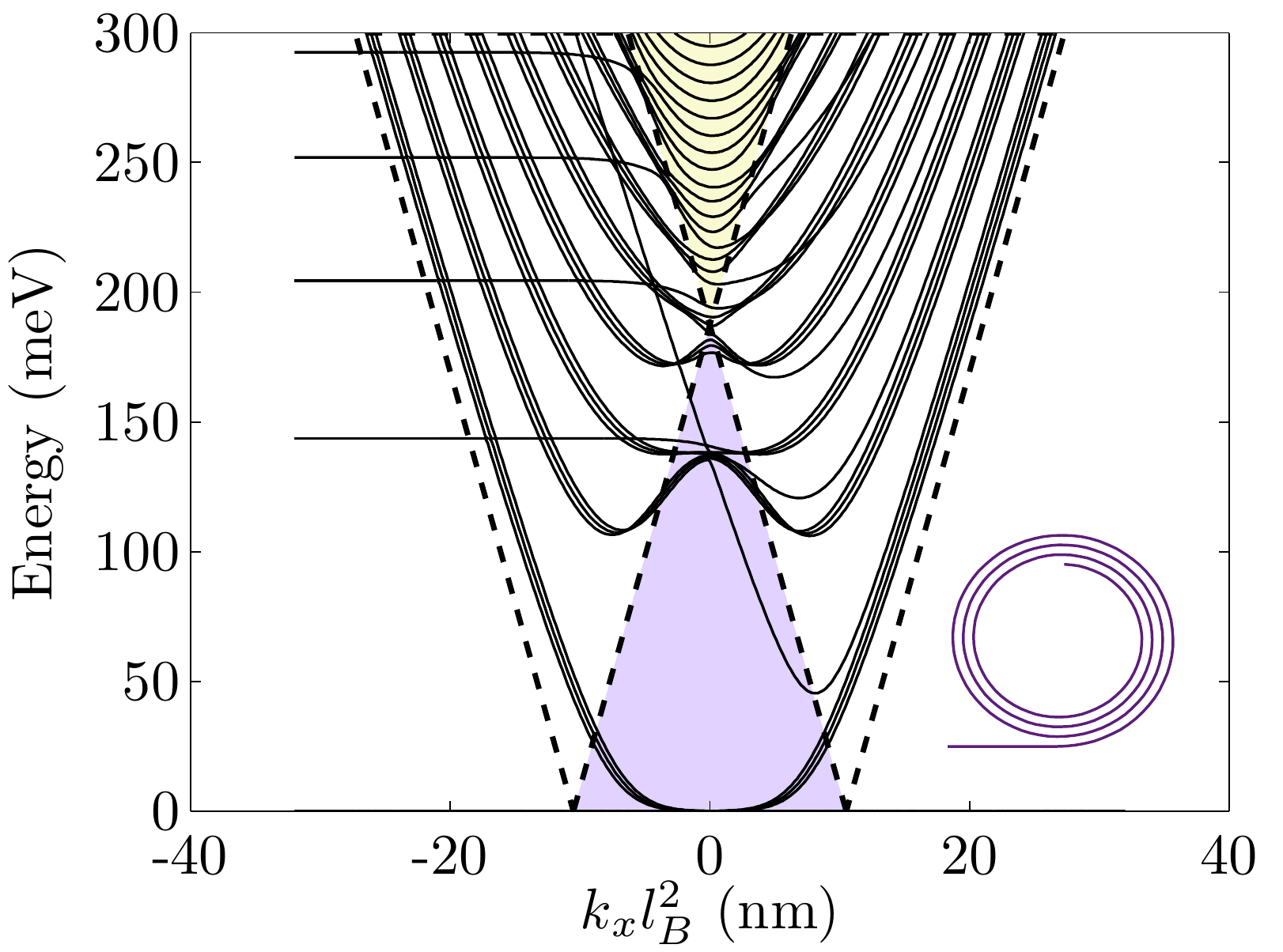}
\caption{\label{fig:Dirac_bands}  (Color online) 
Band structure of a scroll with the inner and outer radii, respectively, $r_i = 9.91$ and $r_o = 11.1\,\text{nm}$ at $B = 20\,\text{T}$ computed from the Dirac equation with the zigzag edge boundary condition.
The dashed lines are defined by equations $E = \hbar v (k_x \pm l_B^{-2} r_o)$. The cross-section of the system is sketched in the bottom right corner of the figure. For clarity, the flat region (full width $100\,\text{nm}$) is truncated and the layer separation inside the scroll is enlarged.
}
\end{center}
\end{figure}
%%%%%%%%%%%%%%%%%%%%%%%%%%%%%%%%%%%%%%%%%%%%%%%%%%%%%%%%%%%%%%%%%%%%%%%%%

One can extend the same approach to other situations, including scrolls clamped and tapered near the contacts~\cite{Meyer2007tso, Fogler2010eoe}. In general, $R_s$ would be augmented by a model-dependent contact resistance, which would suppress the portion of the current going into the scrolls.

Let us now address whether it is justifiable to neglect the effect of the magnetic field on the scrolls. This issue is separate from the role of disorder and contacts. Hence, we can assume for the time that the system is disorder-free and infinite in the $x$-direction. The electron energies $E(k_x)$ can then be labeled by the conserved momentum $k_x$. The low-energy part of the spectrum is easily computed by diagonalizing the effective Dirac Hamiltonian
(suitable for zigzag edges)
\begin{equation}
\mathcal{H} = \hbar v [\sigma_x (k_x - l_B^{-2} Y) - i \sigma_y \partial_y]\,,
\label{eqn:H_Dirac}
\end{equation}
where $v$ is the Fermi velocity, $Y$ is the $y$-coordinate in the original 3D space, $l_B = \sqrt{\hbar c / e B}$ is the magnetic length, $\sigma_\nu$ are the Pauli matrices~\cite{NGPNG09}. In Fig.~\ref{fig:Dirac_bands} we show a representative spectrum for the scroll of outer radius $r_o = 11.1\,\text{nm}$ and arc length $\ell = 100\,\text{nm}$ connected to the flat region of the same width. The discrete energy plateaus come from the flat region of the sample. The rest of the spectrum very much resembles that of a CNT~\cite{Perfetto2007qhe} except the dispersion curves appear in bunches. The literature on the Landau problem for CNT, e.g.,~\cite{Ajiki1993eso, Lee2003sic, Charlier2007eat, Perfetto2007qhe} and other curved 2D systems~\cite{[{For review, see }] Peters2010ctd} readily furnishes an interpretation of these spectral features. The spectrum consists of four regions demarcated approximately along the lines $E = \hbar v (k_x \pm l_B^{-2} r_o)$. These regions  correspond to quasiclassical trajectories of four different types.
The bottom region (shaded purple in Fig.~\ref{fig:Dirac_bands}) can be viewed as the Landau levels of electrons confined near the top of the bottom of the scroll and experiencing the local magnetic field $B_\perp(Y = k_x l_B^2)$. The left and the right regions correspond the snake-like states propagating along the eponymous sides of the scroll. The number of the lines in each dispersion bunch (either three or four in Fig.~\ref{fig:Dirac_bands}) represents the number of layers in the given region of the scroll (top, bottom, left, or right). The energy spacing within these bunches is due to the small difference in the layer radii. If we were to include effects of electron interaction, self-consistent screening of external electric fields, and/or disorder, the near-degeneracy of the bunches would be considerably lifted.

Our crude approximation of neglecting the $B$ field is reasonably accurate inside the top region (shaded gold) of traversing~\cite{Bellucci2010lla}, i.e., spiraling trajectories. Indeed, at $E \gg \hbar v l_B^{-2} r_o$ the
cyclotron radius $R = c E / (e v B_\perp)$ greatly exceeds $r_o$, so that deflection of electrons by the Lorentz force is negligible. The spectrum of the top region is essentially the same as that of a graphene ribbon of width $\ell + \mathcal{O}(l_B)$ at $B = 0$. As $r_o$ decreases, the region of traversing trajectories becomes progressively more dominant. Eventually, at $r_o < l_B$, neglecting the effect of $B$ inside the scrolls is fully justified. In the remainder of the paper we focus on exactly this case.

Dirac Hamiltonian~\eqref{eqn:H_Dirac} may no longer be accurate for very narrow scrolls (note that at, say, $B = 20\,\text{T}$ we have $l_B = 5.7\,\text{nm}$). Therefore, we replace it by the tight-binding model $\mathcal{H} =  \sum_{j} \epsilon_j |j\rangle \langle j| + \gamma_0
\sum_{\langle j, k\rangle} e^{-i \varphi_{j k}} \ket{j} \bra{k}$
defined on a honeycomb lattice of sites $\ket{j}$.
Here $\gamma_{0} = 2.7\,\text{eV}$ is the coupling between nearest neighbors and $\varphi_{j k}$ are the Peierls phases proportional to the magnetic flux through each hexagon of the lattice. This flux is constant in the flat region but oscillates from positive to negative within each turn of the scroll.

For simplicity and numerical efficiency, we work with the rectangular rather than I-shaped sample. To be precise, we consider an armchair nanoribbon composed of 814 dimer lines, 814-aNR. Its width of $100\,\text{nm}$ is significantly smaller than that of suspended graphene flakes studied experimentally. However, this does not qualitatively change any of our main conclusions.

\begin{figure}
\begin{center}
\resizebox{\columnwidth}{!}{\includegraphics{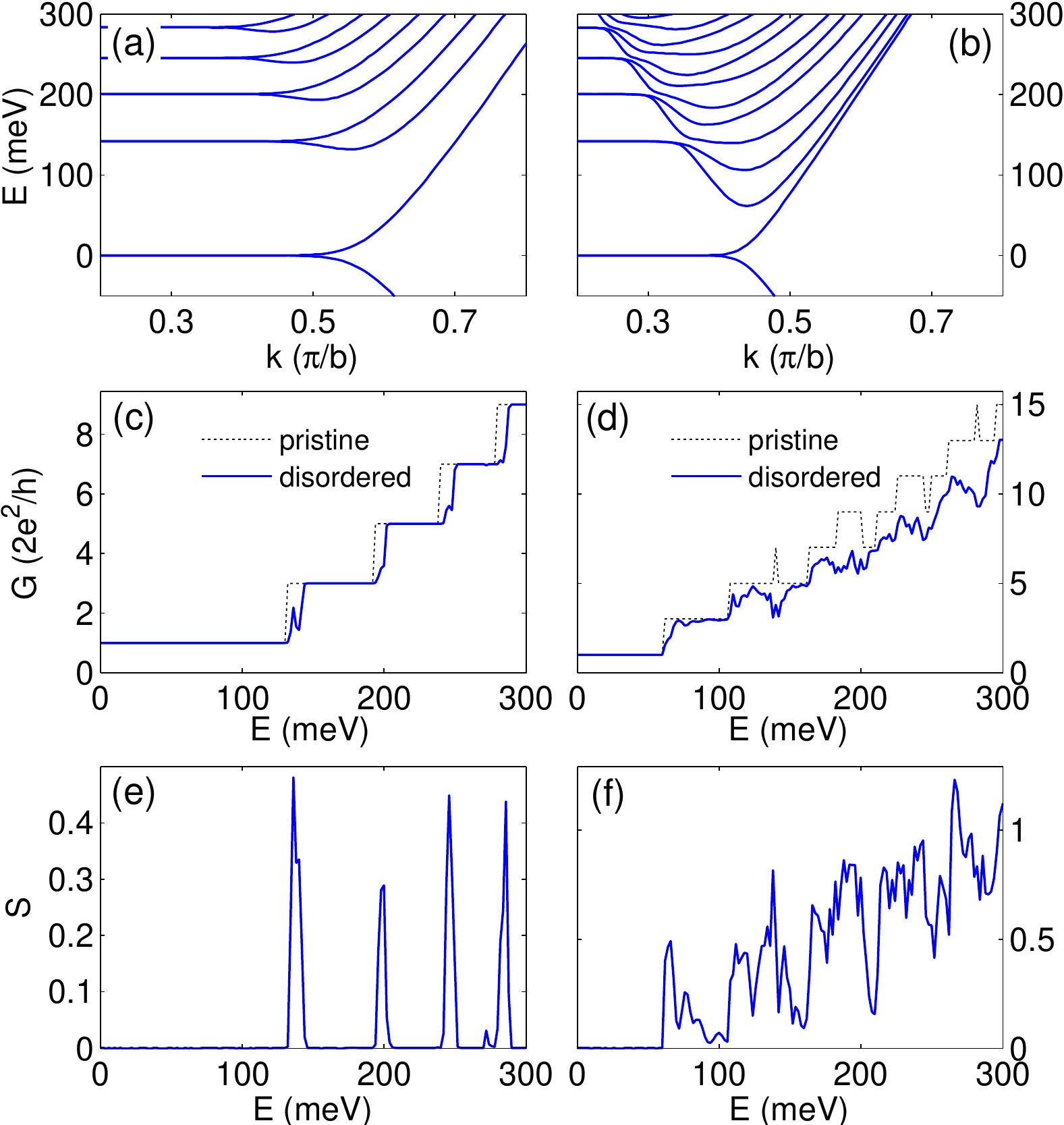}}
\caption{\label{fig:bands}
  (Color online) 
  (a) Band structure of a flat 814-aNR at $B = 20\,\text{T}$.
  (b) The same as (a) in the presence of nanoscrolls of arc length $\ell = 15\,\text{nm}$ and two full turns.
  (c) Conductance as a function of the electron energy for pristine (black dashed lines) and disordered (blue solid lines) 814-aGNRs at $B = 20\,\text{T}$ in the absence of scrolls. 
  (d) The same as (c) in the presence of the scrolls.
  (e) and (f) Shot-noise for a disordered system described in, respectively, (c) and (d).
}
\end{center}
\end{figure}

We start with the disorder-free case where all the on-site energies $\epsilon_j$ are zero. Without the scrolls, we find the familiar sequence of Landau levels in the bulk (low $k$) and the dispersive channels at the edges (high $k$) at $B = 20\,\text{T}$. Except for the narrow energy interval near the bulk levels, the edge channels are chiral. They have the same sign of velocity $d E / d k$, see Fig.~\ref{fig:bands}(a). (The doublet structure of the edge states and the narrow region of nonmonotonic dispersion is a non-generic peculiarity of the armchair edge~\cite{NGPNG09, Ribeiro2011utm}.)

The edge channels undergo a qualitative change when the $\ell = 15\,\text{nm}$ strips at both edges are wrapped into scrolls of two full turns. Both signs of velocity are now present at all $E$, see Fig.~\ref{fig:bands}(b). The energy separation of the edge states is about the same as in the $30\,\text{nm} \approx \ell + 3 l_B$ wide 241-aNR in zero field. Evidently, it has little to do with the bulk Landau level gaps.

%%%%%%%%%%%%%%%%%%%%%%%%%%%%%%%%%%%%%%%%%%%%%%%%%%%%%%%%%%%%%%%%%%%%%%%%%
%%%%%%%%%%%%%%%%%%%%%%%%%%%%%%%%%%%%%%%%%%%%%%%%%%%%%%%%%%%%%%%%%%%%%%%%%

Next we compute differential conductance $G(E) = 1 / R = (2e^2/h) \sum_{m} T_{m}$, where $T_{m}$ are the transmission coefficients for the transmission channels at a given Fermi energy $E$~\cite{[{For computational details, see }] Cresti2008cti}. Only occasionally these channels are of mixed origin. At most $E$, they can be attributed to either the flat or the scrolled regions. Therefore, the earlier formula for resistors connected in parallel, $G = R_f^{-1} + R_s^{-1}$, still applies. However, the magnitude of these resistors is now computed from the quantum expressions.

Since the energy spacing in the scrolled regions in Fig.~\ref{fig:bands}(b) is considerably smaller than the Landau level spacing in the bulk, we deal with the case $R_s \ll R_f$ where the scrolls essentially dominate the transport. This is similar to the large $\rho_{x y} / \rho_{x x}$ limit in Fig.~\ref{fig:current} where the total resistance is $B$-independent. Indeed, let us compare the dashed traces in Figs.~\ref{fig:bands}(c) and (d). Both are perfectly quantized. However, their average slopes differ almost twice. While Fig.~\ref{fig:bands}(c) reflects the quantization of $\rho_{x y}$, Fig.~\ref{fig:bands}(d) basically describes the subband quantization of the 241-aNR in zero field, with not much relation to the actual $B$. Some interplay between the two types of quantization is still present. It leads to, e.g., intriguing nonmonotonic behavior of $G(E)$ in Fig.~\ref{fig:bands}(d), noted previously for other scrolled 2D systems~\cite{Magarill1998bta}.

To make the model more realistic we introduce on-site disorder in a ribbon section of length 210 nm. It includes a short-range disorder generated by randomly varying the on-site energies $\epsilon_j$ within the range $[-25,25]\,\text{meV}$. We also add a finite-range disorder modeled by a sum of 50 Gaussians with range $\xi = 1\,\text{nm}$ centered at random positions and having random strengths in the range $[-500,500]\,\text{meV}$. The disorder causes scattering among counter-propagating channels, which is conveniently quantified by the shot-noise $S = \sum_{m} T_{m} (1-T_{m})$.

When disorder is present, the scroll-free structure exhibits a robust quantized conductance and zero shot-noise, except for energies nearby each onset of new subbands, see the solid line in Fig.~\ref{fig:bands}(c). This situation is typical of the quantum Hall effect. (Some residual backscattering is due to the relatively small width of the system and also due to
the aforementioned aNR-specific nonchiral edge states near the bulk Landau energies.) For the ribbon with scrolled edges, Fig.~\ref{fig:bands}(d), the conductance quantization is considerably degraded with the exception of region corresponding to the first conductance plateau. The enhanced shot-noise, Fig.~\ref{fig:bands}(f), also indicates a stronger backscattering. This behavior is typical of narrow disordered ribbons in which spatial overlap of conduction channels causes significant backscattering even in the presence of a moderate disorder. 

The backscattering of the counter-propagating channels would eventually lead to localization of the nonchiral modes in the scrolls, thus recovering the quantum Hall effect in long ribbons. However, the localization length should scale with the number of channels, i.e., the arc length $\ell$ of the scroll. Therefore, for large enough $\ell$, the localization may not be very important. This can be contrasted with the case of multiwall CNT where the current is confined to the outer shell~\cite{Charlier2007eat}, which reduces the number of channels.

In conclusion, the edge scrolls in suspended graphene have been shown to short-circuit the source-drain current paths in the quantum Hall regime inhibiting the observation of the Hall conductance quantization in such systems.

A.C. acknowledges the Fondation Nanosciences {\it via} the RTRA Dispograph project. 
Funding from MICINN (Spain) through grants FIS2008-00124 and CONSOLIDER CSD2007-00010 is gratefully acknowledged. 
M.F. is supported by UCOP.

%%%%%%%%%%%%%%%%%%%%%%%%%%%%%%%%%%%%%%%%%%%%%%%%%%%%%%%%%%%%%%%%%%%%%%%%%
% \bibliography{ScrolledEdges}
% \end{document}

%Merlin.mbs v4.21 2009-07-09.
%

\end{document}